# The Physics of Strong Magnetic Fields in Neutron Stars[1]

Qiu-he Peng ( qhpeng@nju.edu.cn ), Hao Tong

（Department of Astronomy, Nanjing University, Nanjing, 210093, China）

## Abstract：

In this paper we present a new result: the primal magnetic field $B^{(0)}$ of the collapsed core due to the conservation of magnetic flux during supernova explosion will be greatly boosted up more than 90 times to $B_e^{(in)}$ by the Pauli paramagnetization of the highly degenerate relativistic electron gas just after the formation of the neutron stars. Thus, the observed super-strong magnetic field of neutron stars may originate from the induced Pauli paramagnetization of the highly degenerate relativistic electron gas in the neutron star interiors. Therefore, we have a seemingly more natural explanation of the neutron star surface magnetic field.

## 1. Introduction

It is generally believed that there is a very strong magnetic field, $B > 10^{12}$ Gauss, for most of the neutron stars (e.g. Shapiro & Teukolski, 1984). There probably are magnetars with ultra-strong magnetic field strength exceed the quantum critical threshold, $H_{cr} = 4.414 \times 10^{13}$ Gauss (Duncan & Thompson 1992; Paczynski, 1992; Usov, 1992; Thompson & Duncan, 1995, 1996). Anomalous X-ray Pulsars (AXPs) and Soft Gamma Repeaters (SGRs) are classes of candidates to magnetars (e.g. Kouveliotou et al., 1998, 1999; Hurley et al., 1999; Mereghetti & Stella, 1995; Wilson et al., 1999; Kaspi, Chakrabarty & Steinberger, 1999). The magnetic field of the magnetars may be so strong as to reach two orders of magnitude above the quantum critical threshold, $10^{14}$-$10^{15}$ Gauss.

What is the origin of the strong magnetic field of some neutron stars with the magnetic field such as near the quantum critical threshold $H_{cr}$? What is the origin of the ultra-strong magnetic field of the magnetars? They are indeed very interesting questions.

Generally, it is believed that the strong magnetic field of neutron stars originated from the collapse of the core of a supernova with the conservation of magnetic flux. The initial magnetic field strength $B^{(0)}$ after the collapse may be estimated as

$$B^{(0)} = (\frac{R_{Core}}{R_{NS}})^2 B_* \sim (10^8 - 10^{10}) B_* \text{ Gauss} \qquad (1)$$

where $B_*$ is the magnetic field strength of the presupernova core with radius $R_{Core}$. As typical values, we have taken $R_{Core} \sim (10^5 - 10^6)$ km and $R_{NS} \sim 10$ km. It is easily seen that the magnetic field of the collapsed supernova core is still much less than $10^{13}$ Gauss, even for a very strong

[1] This research is supported by Chinese National Science Foundation No.10573011, No.10273006, and the Doctoral Program Foundation of State Education Commission of China

magnetic field of the presupernova, $B_* \sim 10^3$ Gauss. This means that it is rather difficult to get the strong magnetic field such as those expected for some of the pulsars with $B > 10^{13}$ Gauss, especially for the ultra-strong field strength of the magnetars, by the collapse process alone.

In this paper we will address the first question above in detail. We are going to calculate the strong magnetic field produced by the Pauli paramagnetic moment of highly degenerate Fermi (neutron, proton or electron) system. As to the second question of the formation of the magnetars, we are going to discuss it in detail in another paper (Peng, 2006; Peng, Luo & Chou, 2006). We have found the solution that the ultra-strong magnetic fields of the magnetars are produced by the induced magnetic moment of the $^3P_2$ neutron pairs in the anisotropic neutron superfluid when the interior temperature of the neutron stars decrease down to much lower than $10^7$ K.

Treatment of Pauli paramagnetism in terrestrial case has become a paradigm of Fermi statistics. Employing the grand canonical formalism, starting from the partition function, we can consider the Pauli paramagnetism and Landau diamagnetism in a unified way. We will follow the terrestrial formalism in our celestial case (earliest treatment see Canuto & Chiu, 1968). Of course there are other versions (Mandal & Chakrabarty, 2002). Naturally, the main results should be and are consistent with each other.

## 2. The magnetic field produced by the Pauli paramagnetic moment of a highly degenerate relativistic electron system

The spin of an electron is depicted by $\vec{S} = \frac{1}{2}\hbar\vec{\sigma}$, $|\sigma| = 1$, $\sigma_z = +1, -1$. The electron possesses a magnetic moment, $\vec{\mu} = -\mu_e \vec{\sigma}$, $\mu_e = 0.927 \times 10^{-20} erg/G$. Where $\mu_e$ is the Bohr magneton. The energy of an electron in an external magnetic field B is: $\sigma\mu_e B$. The induced magnetic moment (or Pauli paramagnetization) of a Fermi system may be calculated by the standard method of quantum statistics.

$$\mu^{(in)} = k_B T \frac{\partial \ln \Xi}{\partial B} \tag{2}$$

$$\ln \Xi = \sum_{\sigma=\pm 1} \int_0^\infty N(\varepsilon) \ln[1 + \exp\{\beta(\psi - \varepsilon_k - \sigma\mu_e B)\}] d\varepsilon \tag{3}$$

where $\beta = 1/k_B T$, $\Xi$ is the grand partition function of the system, B is the external magnetic field, $k_B$ is the Boltzmann's constant and $\Psi$ the chemical potential of the Fermi gas (the Fermi energy $E_F$). $N(\varepsilon)$ is the density of states, $N(\varepsilon)d\varepsilon = \frac{4\pi V}{h^3} p^2 dp$. The Fermi sphere of the electron gas is spherically symmetric when the applied magnetic field is much weaker than the Landau critical value ($B_{cr} = 4.414 \times 10^{13}$ Gauss).

In the regime $\mu_e B(<1MeV) << E_F(e)(>60MeV)$, the integrand may be expanded in a series of $\sigma\mu_e B$ as follows

$$\ln[1+\exp\{\beta(\psi-\varepsilon-\sigma\mu_e B)\}] = \ln[1+e^{\beta(\psi-\varepsilon)}] - \beta\sigma\mu_e B\bar{n}(\varepsilon) + \frac{1}{2}(\beta\sigma\mu_e B)^2 \bar{n}(\varepsilon)[1-\bar{n}(\varepsilon)] \quad (4)$$

Where $\bar{n}(\varepsilon)$ is the average occupation number of the electrons at the quantum state with energy ε,

$$\bar{n}(\varepsilon) = [1+e^{\beta(\varepsilon-\psi)}]^{-1} \quad (5)$$

The sum of the second term for σ(=-1, +1) equals to zero and both the sum of the first term and third term for σ(=-1, +1) are just to be doubled when eq.(4) is substituted into eq.(3).

$$\ln\Xi = 2\int_0^\infty d\varepsilon N(\varepsilon)\ln[1+e^{\beta(\psi-\varepsilon)}] + (\beta\mu_e B)^2 \int_0^\infty d\varepsilon N(\varepsilon)\bar{n}(\varepsilon)[1-\bar{n}(\varepsilon)] \quad (6)$$

The first term has no contribution for calculation of the magnetic moment since it is independent of the magnetic field. Using the relations

$$\bar{n}(\varepsilon)[1-\bar{n}(\varepsilon)] = -kT\frac{d\bar{n}(\varepsilon)}{d\varepsilon} \quad (7)$$

$$-\int_0^\infty d\varepsilon N(\varepsilon)\frac{d\bar{n}(\varepsilon)}{d\varepsilon} = N(\psi) + \frac{\pi^2}{6}(kT)^2 \frac{d^2 N(\psi)}{d\psi^2} \quad (8)$$

we may calculate the induced Pauli paramagnetic moment for the electron system as follows:

$$\mu^{(in)} = kT\frac{\partial \ln\Xi}{\partial B} = 2\mu_e^2 B^{(0)} N(\psi)[1+\frac{\pi^2}{6}(kT)^2 \frac{1}{N(\psi)}\frac{d^2 N(\psi)}{d\psi^2}]. \quad (9)$$

Here $B^{(0)}$ is the background magnetic field.

The electron gas in the neutron stars is highly relativistic and degenerate. The relation between energy and momentum is $\varepsilon = cp$. We then have

$$N(\varepsilon) = \frac{4\pi V}{(hc)^3}\varepsilon^2 \quad (10)$$

$$\frac{d^2 N(\psi)/d\psi^2}{N(\psi)} = \frac{2}{E_F^2} \quad (11)$$

$$\mu^{(in)} = 2\mu_e^2 B^{(0)} N(\psi)\{1+\frac{\pi^2}{3}(\frac{kT}{E_F(e)})^2\} \approx 2\mu_e^2 B^{(0)} N(\psi) \quad (12)$$

(Same as the non-relativistic case. See also Feng & Jin's deduction.)

Employing the dipolar model of neutron stars, it can be estimated that the induced magnetic field due to the Pauli paramagnetism of the electron system is

$$B_e^{(in)} = \frac{2\mu^{(in)}}{R_{NS}^3} \approx AB^{(0)} \quad (13)$$

$$A = \frac{4\mu_e^2}{R_{NS}^3} N(E_F(e)) \tag{14}$$

Or we may express the amplification factor of the magnetic field by combining eq.(10) for the highly relativistic degenerate electron gas in the form

$$A \approx \frac{64\pi^2}{3} \frac{\mu_e^2}{(hc)^3} E_F^2(e) \tag{15}$$

Using the relations $n_e = \frac{8\pi}{3h^3} p_F^3$, $p_F = E_F/c$, $n_e = Y_e N_A \rho$, we finally obtain

$$\begin{aligned} A &= \frac{64\pi^2}{3} (\frac{3}{8\pi})^{2/3} \frac{\mu_e^2}{hc} N_A^{2/3} (Y_e \rho)^{2/3} \\ &\approx 0.91 \times 10^2 [\frac{Y_e}{0.05} \frac{\rho}{\rho_{nuc}}]^{2/3} \end{aligned} \tag{16}$$

Where $Y_e$ is the mean number of electrons per baryon. In addition, an electron gas also has a diamagnetic susceptibility originating from changing the orbital states by the applied magnetic field, which is called Landau diamagnetic susceptibility. Besides the Pauli paramagnetization, therefore, the Landau diamagnetization of the electron gas must also be included.

For the non-relativistic but highly degenerate electron gas (terrestrial laboratory case), the Landau diamagnetic susceptibility is $-\frac{1}{3}$ the Pauli paramagnetic susceptibility (Feng and Jin, 2005, for detailed caculation see Pathria, 2003). And the Landau diamagnetic susceptibility is much smaller than the Pauli paramagnetic susceptibility for relativistic highly degenerate electron gas, about $-10^{-4}$, see the appendix for detailed discussion.

It may be deduced that the induced magnetic field produced by the Pauli paramagnetization for the highly degenerate relativistic electron gas in neutron stars is 90 times stronger than the applied magnetic field, which is just the initial magnetic field $B^{(0)}$ due to the gravitational collapse of the presupernova core

As to the neutron or the proton system, their Pauli paramagnetization is much weaker than that of the electron system in neutron stars. The main reason is:
1. the inherent magnetic moment of a neutron or a proton is only one in a thousand of the electron magnetic moment.

Pauli paramagnetization of the neutron system may be obtained simply by calculating the amplification factor, A, of the magnetic field by eq. (14) with the substitution $\mu_e \to \mu_n$ and using the expression of the energy level density at the Fermi surface of the non-relativistic degenerate neutron system $N(\varepsilon) = \frac{V}{2\pi^2 \hbar^3} (2m_n)^{1/2} m_n \varepsilon^{1/2}$. Taking V= $(4\pi/3)R^3$, R≈$R_{NS}$, we may get A ~ $2\times 10^{-3}$. So the Pauli paramagnetic magnetization of the neutron and the proton system may be neglected.

According to Mandal & Chakrabarty (2002), under the magnetic field of neutron stars, the neutron or proton system never becomes fully polarized. While to that of electron, it is very close to fully polarized. Qualitatively, it is consistent with our result here. The difference is we have made quantitative calculations.

Therefore, we conclude that the observed strong magnetic field of neutron stars may originate from the induced magnetic field produced by the Pauli paramagnetization of the highly degenerate relativistic electron gas in the neutron star interiors.

## 3. Induced Pauli paramagnetization of a highly degenerate relativistic electron gas under Landau critical magnetic field

We will now consider the Landau quantum effect on the motion of the electrons in the direction perpendicular to the applied magnetic field. The energy of an electron in a homogeneous magnetic field B is

$$E^2(p_z, B, n, \sigma) = m^2c^4 + p_z^2c^2 + (2n+1+\sigma)2mc^2\mu_e B \tag{17}$$

Comparing with the relativistic energy expression

$$E^2 = m^2c^4 + p_z^2c^2 + p_\perp^2c^2 \tag{18}$$

We may get

$$p_\perp^2(B, n, \sigma) = 2m(2n+1+\sigma)\mu_e B \tag{19}$$

Where $\sigma$ is the projection of the spin quantum number, $\sigma = \pm 1$, $n$ is the quantum number of the Landau energy level, $n = 0,1,2,3\ldots$.

For the electron gas in the neutron stars, $E_F(e) \sim 60\text{-}100\ MeV$, $mc^2 = 0.511 MeV$, $\mu_e B \sim 0.3\ (B/B_{cr})$ $MeV$ ($B_{cr} = 4.414 \times 10^{13}$ Gauss), we then have

$$E_F(e) \approx p_z c + \varepsilon \quad , \quad \varepsilon \ll p_z c \tag{20}$$

$$\varepsilon = \frac{1}{2p_z c}[m^2c^4 + (2n+1+\sigma)2mc^2\mu_e B] \tag{21}$$

The Fermi sphere with spherical symmetry for the electron system will be distorted by the applied strong magnetic field into a Landau cylinder. The (level) density of state for the Landau cylinder is

$$N(\varepsilon)d\varepsilon = \frac{V}{h^3}\pi p_\perp^2 \bigg|_n^{n+1} dp_z = \frac{4\pi V}{h^3} m\mu_e B dp_z \tag{22}$$

or

$$N(B, n, \sigma) = \frac{4\pi V}{h^3 c} m\mu_e B \tag{23}$$

due to $d\varepsilon = cdp_z$. The total level density of energy is

$$N(\varepsilon, B, T) = \sum_{n,\sigma=\pm 1} N(B, n, \sigma) \tag{24}$$

Near the Fermi surface,

$$N(E_F, B, T) = \frac{4\pi V m \mu_e B}{h^3 c} G, \tag{25}$$

$$G = \sum_{n, \sigma = \pm 1} 1 \tag{26}$$

The great majority of the electrons is in the lower energy states $n = 0,1,2\ldots$ and then G is less than 10 generally. We may estimate the value of the amplification factor of the magnetic field, $A$, in eq.(13) by using eq.(14) and eq.(26),

$$A = \frac{64\pi^2}{3} \frac{m\mu_e^3 B^{(0)}}{h^3 c} G = 7.7 \times 10^{-4} (\frac{B^{(0)}}{B_{cr}}) G \sim 10^{-2} \tag{28}$$

Therefore we may conclude that the induced Pauli paramagnetization of the highly degenerate relativistic electron gas under the influence of the Landau critical magnetic field is very weak and may be neglected. This is because the level density of energy of the Landau cylinder is much lower (about $10^{-4}$) than that of the Fermi surface with spherical symmetry in a weak magnetic field(see earliest discussion by Canuto and Chiu, 1968).

Consequently, the ultra-strong magnetic field, $B \sim 10^{14}$ -$10^{15}$ Gauss of the magnetars cannot originates from the induced Pauli paramagnetization of the highly degenerate relativistic electron gas under the Landau critical magnetic field. We have to find other novel model for the origin of the ultra-strong magnetic field of the magnetars ((Peng, 2006; Peng, Luo & Choul, 2006).

## 4. Conclusions

The primal magnetic field will be shortly boosted up to about 90 times by the Pauli paramagnetization of the highly degenerate relativistic electron gas in the neutron star interior. This mechanism naturally explains the neutron star strong surface magnetic field. But in the case of magnetars, other models should be employed.

For the main sequence stars, the $A_p$ stars have the strongest surface magnetic field about 1000 Gauss. Then, we obtain the upper limit for neutron star magnetic field about $10^{15}$ Gauss. But we should bear in mind that the $A_p$ stars is not the progenitors of supernova.

Our upper limit here can be compared with Mandal & Chakrabarty (2002). They have obtained an upper limit of $10^{16}$ Gauss, which is the saturation value for a magnetic dipole system.

## Acknowledgment

Author A is very grateful to Professor Chich-gang Chou for his help of improving English of the paper.

## Appendix: Landau diamagnetism in the relativistic case

The energy momentum relation is given by the Landau solution of the Dirac equation in the presence of a homogeneous magnetic field

$$E^2 = m^2c^4 + p_z^2 c^2 + (2n+1+\sigma)\hbar ceB, \tag{A1}$$

where n is orbital quantum number having values 0, 1 ,2 and so on, $\sigma$ is polarization quantum number, as stated in the beginning of section two. The upper equation can be modified employing Bohr magneton $\mu_B = \dfrac{e\hbar}{2mc}$, thus we can rewrite the energy momentum relation as follows

$$E^2 = m^2c^4 + p_z^2 c^2 + (2n+1+\sigma)2mc^2\mu_B B. \tag{A2}$$

Following the traditional treatment of Landau diamagnetism, we throw away the Zeeman energy term

$$E^2 = m^2c^4 + p_z^2 c^2 + (2n+1)2mc^2\mu_B B, \tag{A3}$$

and henceforth the energy can be treated as a function of $n+\dfrac{1}{2}$.

After the energy momentum relation is obtained, we can calculate the multiplicity factor

$$\frac{1}{h^2}\int dp_x dp_y = \frac{1}{h^2}\pi p_\perp^2 \Big|_n^{n+1} = \frac{4\pi m\mu_B B}{h^2}, \tag{A4}$$

which is the same as in the nonrelativistic case(Pathria, 2003). This will be usefull in calculating the grand partition function.

In our case, the grand partition function is

$$\Xi = \sum_\varepsilon \log(1+ze^{-\beta\varepsilon}) = 2\int \frac{dp_z}{h} \sum_{n=0}^{\infty} \frac{4\pi m\mu_B B}{h^2} \log(1+ze^{-\beta\varepsilon(n+\frac{1}{2})})$$
$$= \frac{8\pi m\mu_B B}{h^3} \int_{-\infty}^{\infty} dp_z \sum_{n=0}^{\infty} \log(1+ze^{-\beta\varepsilon(n+\frac{1}{2})}),$$
(A5)

where z is fugacity of the system. As pointed out above, $\varepsilon$ is an even function of the z-direction momentum, and with the help of the Euler summation formula

$$\sum_{j=0}^{\infty} f(j+\frac{1}{2}) \cong \int_0^{\infty} f(x)dx + \frac{1}{24} f'(0),$$
(A6)

the grand partition function now becomes

$$\Xi = \frac{8\pi m\mu_B B}{h^3} 2\int_0^{+\infty} dp_z \frac{1}{24} \log(1+ze^{-\beta\varepsilon(n+\frac{1}{2})})',$$
(A7)

where a prime means derivation with respect to the variable $n+\frac{1}{2}$ in the expression of the $\varepsilon$, and we have drop out terms independent of B.

In a magnetized relativistic gas, the magnetic moment is $\mu = \frac{1}{\beta} \frac{\partial \ln \Xi}{\partial B}$, it has analytical expression if we adopt the zero temperature approximation

$$\mu = -\frac{1}{3} \frac{1}{x_F^2} \sinh^{-1}(x_F) \, 2\mu_B^2 g(\varepsilon_F) B,$$
(A8)

where $x_F$ is the dimensionless Fermi momentum $x_F = \frac{p_F}{mc}$, $g(\varepsilon_F)$ is the density of states at the Fermi surface. Compare this with the Pauli paramagnetism(eq.12), we obtain

$$\mu = -\frac{1}{3} \frac{1}{x_F^2} \sinh^{-1}(x_F) \mu_{Pauli}$$
$$= -10^{-4} \mu_{Pauli.}$$
(A9)

where $x_F \sim 200$, as in the case of neutron stars(Shapiro, p310).

That is, in the relativistic case, the Landau diamagnetism is only $-10^{-4}$ the Pauli paramagnetism, thus insignificant.